\documentclass[a4paper,10pt]{article}
\usepackage[utf8]{inputenc}
\usepackage{amssymb}
\usepackage{amsmath}
\usepackage{arydshln}
\usepackage{geometry}
\usepackage{hyperref}
\hypersetup{
  colorlinks   = true, 
  urlcolor     = blue, 
  linkcolor    = blue, 
  citecolor   = red 
}

\begin{document}

\begin{titlepage}
 
 \begin{center}
 \hfill
 \\	[22mm]
 
  {\Huge{Double Soft Limit of Graviton Amplitude from the Cachazo-He-Yuan Formalism}}\\
  \vspace{16mm}
  
\large {Arnab Priya Saha}\\
\vspace{5mm}
{ \it Institute of Mathematical Sciences, \\ }
{ \it C.I.T Campus, Taramani, Chennai 600113, India\\}
and\\
{ \it Homi Bhabha National Institute, \\}
{ \it Training School Complex, Anushakti Nagar, Mumbai 400085, India.\\ }
\vspace{2mm}
email: \href{mailto: arnabps@imsc.res.in}{arnabps@imsc.res.in}

\end{center}

\medskip
\vspace{40mm}

\begin{abstract}

We present a complete analysis for double soft limit of graviton scattering amplitude using the formalism proposed by Cachazo, He and Yuan. Our results agree with that obtained via BCFW recursion relations in \cite{Klose}. In addition we find precise relations between degenerate and nondegenerate solutions of scattering equations with local and nonlocal terms in the soft factor.    

\end{abstract}

\end{titlepage}

\section{Introduction}
In a series of seminal papers \cite{PRL, Cachazo:KLT, Cachazo:graviton, Cachazo:YM, Yuan} Cachazo, He and Yuan developed a remarkable formalism of calculating scattering amplitude for a large variety of theories like gravity, Yang-Mills, non-linear sigma models, Dirac-Born-Infeld and many others using integrals over moduli space of punctured Riemann spheres. They showed for scattering of $n$ massless particles tree-level scattering amplitude is given by 

\begin{equation}\label{scat-amplitude}
 M_{n} = \int \frac{\mathrm{d}\sigma_{1}\cdots\mathrm{d}\sigma_{n}}{\mathrm{vol}\mathbb{SL}(2,\mathbb{C})} \sideset{}{'}\prod\limits_{a} \delta \left( \sum\limits_{b \ne a}\frac{k_{a}.k_{b}}
		      {\sigma_{a}-\sigma_{b}}\right) I_{n}(\{k,\epsilon,\sigma\}).
\end{equation}
$\sigma$s are the coordinates of punctures on the complex spheres and the integrand $I_{n}$ is theory dependent. Arguments of the delta functions are the so called scattering equations given by

\begin{equation}
 f_{a} := \sum\limits_{\substack{b=1{}\\b\ne a}}^{n} \frac{k_{a}.k_{b}}{\sigma_{a}-\sigma_{b}} = 0, \qquad a\in\{1,2,\cdots ,n\}.
\end{equation}
These scattering equations are invariant under $\mathbb{SL}(2,\mathbb{C})$ transformations which implies we can fix any three out of the $n$ $\sigma$s to arbitrary values like $0, 1$ and $\infty$. Division of the measure by vol$\mathbb{SL}(2,\mathbb{C})$ is required to fix the $\mathbb{SL}(2,\mathbb{C})$ gauge redundancy. Because of the presence of delta functions the integrals are actually localized to the $(n-3)!$ solutions of the $n$ scattering equations. An important aspect of CHY representations is they hold in arbitrary dimensions. One loop scattering amplitudes from Riemann sphere for supergravity, super-Yang-Mills and non-supersymmetric cases have been derived in \cite{GeyerPRL, Geyer:oneloop}. One loop corrections to scattering amplitudes of scalars and gauge bosons have been done in \cite{He:loop, CHY:loop}. In \cite{Geyer:twoloop} amplitudes for supergravity and super-Yang-Mills at two loop have been presented. 

An outstanding outcome of CHY formalism is the extensive study of soft theorems in a large class of theories \cite{Cachazo:softtheoryext}. Single soft limit as well as multiple soft limits of scattering amplitudes can be obtained with remarkable simplicity in this method \cite{PRL, DoubleSoftPRD}. Subleading soft factors for Yang-Mills and gravity amplitudes have been derived in \cite{Schwab, Afkhami, Zlotnikov, Kalousios} in CHY formalism. These subleading soft factors were first derived in \cite{CachazoStrominger} using BCFW recursion relations. Elaborate studies on the factorization properties of graviton amplitudes can be found in \cite{White, Broedel}. For gravity one can write

\begin{equation}
 M_{n+1}\left(k_{1}, k_{2}, \cdots k_{n}, \tau q\right) \xrightarrow{\tau\rightarrow 0} \left(S^{(0)}(q) + S^{(1)}(q) + S^{(2)}(q)\right)M_{n}\left(k_{1}, k_{2}, \cdots k_{n}\right) + \mathcal{O}(\tau^{2}),
\end{equation}
where the soft factors are given by

\begin{eqnarray}\label{soft factors}
 S^{(0)}(q) & = & \frac{1}{\tau}\sum\limits_{a=1}^{n}\frac{\epsilon_{\mu\nu} k_{a}^{\mu}k_{a}^{\nu}}{q\cdot k_{a}}\nonumber\\
 S^{(1)}(q) & = & \sum\limits_{a=1}^{n} \frac{\epsilon_{\mu\nu}k_{a}^{\mu}q_{\rho}\hat{J}^{\rho\nu}_{a}}{q\cdot k_{a}} \nonumber\\
 S^{(2)}(q) & = & \frac{\tau}{2} \sum\limits_{a=1}^{n} \frac{\epsilon_{\mu\nu}q_{\rho}q_{\sigma}\hat{J}^{\rho\mu}_{a}\hat{J}^{\sigma\nu}_{a}}{q\cdot k_{a}}.
\end{eqnarray}

Following the method depicted in \cite{DoubleSoftPRD} we initiated the study of double soft limit of tree level gravity scattering amplitude in \cite{Saha} and the result thus obtained was matched with Feynman diagrammatic. However we observed two important features of our result:  

\begin{itemize}
 \item The derived soft factor was of $\mathcal{O}(\tau^{-1})$ whereas from Feynman diagrammatic we found a term of $\mathcal{O}(\tau^{-2})$. Naively looking at expressions (\ref{soft factors}) it is justified to expect a leading double soft factor of $\mathcal{O}(\tau^{-2})$.
 
 \item We could only obtain the  term which corresponds to the local processes where two soft gravitons are emitted from single hard legs. Double soft factors of gravity have been derived using BCFW analysis in \cite{Klose}. There for simultaneous double soft emissions the leading and subleading factors are found out to be

\begin{eqnarray}\label{BCFW soft factors}
 S^{(0)}(1^{h_{1}},2^{h_{2}}) & = & S^{(0)}(1^{h_{1}})S^{(0)}(2^{h_{2}}) \nonumber\\
 S^{(1)}(1^{h_{1}},2^{h_{2}}) & = & S^{(0)}(1^{h_{1}})S^{(1)}(2^{h_{2}}) + S^{(0)}(2^{h_{2}})S^{(1)}(1^{h_{1}}) + S^{(1)}(1^{h_{1}},2^{h_{2}})\mid_{\text{c}},
\end{eqnarray}
where $S^{(i)}(q^{h_{m}})$ is the single soft factor corresponding to $m$-th soft particle with helicity $h_{m}$ at $i$-th order. The result of \cite{Saha} precisely matches with the last subleading term which is also called the contact term.
\end{itemize}
The aim of this study is to address the above mentioned issues. It turns out that we need to extend the formalism set up in (\cite{DoubleSoftPRD}) and consider more terms (which are called non-degenerate contributions) to complete the analysis. In this paper we include this additional contributions to derive the so called noncontact terms of (\ref{BCFW soft factors}), which can be written as products of two single soft factors. Multiple soft limits of gluons, gravitons and super-Yang-Mills theories have been discussed in \cite{VolovichZlotnikov} where the authors showed leading order multiple-soft factor for graviton is given by the product of multiple single-soft factors.

This paper is organized as follows: In Sec.(\ref{sec:CHYDS}) we review the double soft limit of scattering amplitude in CHY representation. We give a detail discussion of degenerate and nondegenerate solutions. We derive double soft factors at leading and subleading orders for gravity amplitude in Sec.(\ref{sec:DS gravity}) and show our result matches with that of \cite{Klose}. We discuss the important aspects of our results in Sec.(\ref{sec:discussion}) before ending with some concluding remarks and future directions of work in Sec.(\ref{sec:conclusion}).

\section{Double Soft in CHY}
\label{sec:CHYDS}

Scattering amplitude for $n$ massless particles is given in CHY representation by Eq.(\ref{scat-amplitude}). Our starting point is $n+2$ point amplitude which can be written as

\begin{equation}\label{int rep}
 M_{n+2} = \int \frac{\mathrm{d}^{n}\sigma \mathrm{d}\sigma_{n+1}\mathrm{d}\sigma_{n+2}}{\mathrm{vol}\mathbb{SL}(2,\mathbb{C})} \sideset{}{'}\prod\limits_{a=1}^{n} \delta \left( f_{a}\right)\delta\left(f_{n+1}\right)\delta\left(f_{n+2}\right) I_{n+2}(\{k,\epsilon,\sigma\}).
\end{equation}
Here we have separated out $(n+1)$th and $(n+2)$th integrals for reasons that will be clear shortly. In the soft limits where two of the particles' momenta tend to zero at the same rate (let us denote the soft momenta by $k_{n+1} = \tau p$ and $k_{n+2} = \tau q$ with $\tau \rightarrow 0$) the scattering equations $f_{\alpha}$ can be written as 

\begin{equation}\label{scattering eqn}
 f_{\alpha} = \begin{cases}
          \sum\limits_{\substack{b=1{}\\b\ne a}}^{n} \frac{k_{a}.k_{b}}{\sigma_{a}-\sigma_{b}} + \frac{\tau k_{a}.p}{\sigma_{a}-\sigma_{n+1}} 
          + \frac{\tau k_{a}.q}{\sigma_{a}-\sigma_{n+2}}, & \alpha \in \{1,2,\cdots n\}\\
          \sum\limits_{b=1}^{n} \frac{\tau k_{b}.p}{\sigma_{n+1}-\sigma_{b}} + \frac{\tau^{2}p.q}{\sigma_{n+1}-\sigma_{n+2}}, & \alpha = n+1\\
          \sum\limits_{b=1}^{n} \frac{\tau k_{b}.p}{\sigma_{n+2}-\sigma_{b}} - \frac{\tau^{2}p.q}{\sigma_{n+1}-\sigma_{n+2}}, & \alpha = n+2.
         \end{cases}
\end{equation}
Thus effectively in the vanishing limit of $\tau$ there are $n$ scattering equations for $n$ hard particles. Solving these $n$ equations one obtains solutions for $\sigma_{1}, \sigma_{2},\cdots, \sigma_{n}$ and there are $(n-3)!$ such solution sets. The last two scattering equations do not provide any solution, rather they are used to transform the $\sigma_{n+1}$ and $\sigma_{n+2}$ integrals to contour integrals. As we will see there are subtleties in performing these contour integrals depending on the behavior of $|\sigma_{n+1}-\sigma_{n+2}|$. In the seminal paper \cite{DoubleSoftPRD} the authors classified the behavior in two categories: 1) \textit{non-degenerate solutions} - when $|\sigma_{n+1}-\sigma_{n+2}| \sim \mathcal{O}(\tau^{0})$ and 2) \textit{degenerate solutions} - when $|\sigma_{n+1}-\sigma_{n+2}| \sim \mathcal{O}(\tau)$. It was shown that for theories like sGal, DBI, EMS, NLSM and YMS the leading order contribution come from the degenerate one. However in case of pure gravity (which is given by Einstein-Hilbert action) we find the opposite feature, non-degenerate contribution dominates over the degenerate one. We will elaborate on this issue in more details as we proceed.

\subsection{Non-degenerate Case}

Here we consider the situation when $|\sigma_{n+1} - \sigma_{n+2}|\sim \mathcal{O}\left(\tau^{0}\right)$. This implies that the two soft punctures never overlap each other. The delta functions corresponding to the last two scattering equations, $f_{n+1}$ and $f_{n+2}$ now transform the integrations of $\sigma_{n+1}$ and $\sigma_{n+2}$ variables to independent contour integrals where each of $\sigma_{n+1}$ and $\sigma_{n+2}$ wraps over solutions of $\sigma_{a},\: \forall a\in \{1,2,\cdots, n\}$. Clearly this gives a hint of nonlocal processes where soft particles can be emitted from different external hard particles. In this case the scattering amplitude (\ref{int rep}) takes the form

\begin{equation}\label{M-expansion}
 M_{n+2} \rightarrow \int \frac{\mathrm{d}^{n}\sigma}{\mathrm{vol}\mathbb{SL}(2,\mathbb{C})}\left[\sum\limits_{i=1}^{n}\oint\limits_{|\sigma_{n+1}-\sigma_{i}|\rightarrow 0}\frac{\mathrm{d}\sigma_{n+1}}{f_{n+1}}\right]\left[\sum\limits_{j=1}^{n}\oint\limits_{|\sigma_{n+2}-\sigma_{j}|\rightarrow 0}\frac{\mathrm{d}\sigma_{n+2}}{f_{n+2}}\right] \sideset{}{'}\prod\limits_{a=1}^{n} \delta \left(f_{a}\right) I_{n+2}(\{k,\epsilon,\sigma\}).
\end{equation}
From here onwards we will drop the summation signs and assume sum over the contour integrals is implied. Both the measure and the integrand can be expanded in orders of $\tau$ parameter as follows. 

\begin{eqnarray}\label{f-expansion}
 \frac{1}{f_{n+1}}\frac{1}{f_{n+2}} & = & \frac{1}{\tau^{2}} \frac{1}{\sum\limits_{a}\frac{p\cdot k_{a}}{\sigma_{n+1}-\sigma_{a}}}\left[1 - \tau\frac{\frac{p\cdot q}{\sigma_{n+1}-\sigma_{n+2}}}{\sum\limits_{a'}\frac{p\cdot k_{a'}}{\sigma_{n+1}-\sigma_{a'}}} + \cdots \right] \frac{1}{\sum\limits_{b}\frac{q\cdot k_{b}}{\sigma_{n+2}-\sigma_{b}}}\left[1+\tau\frac{\frac{p\cdot q}{\sigma_{n+1}-\sigma_{n+2}}}{\sum\limits_{b'}\frac{q\cdot k_{b'}}{\sigma_{n+2}-\sigma_{b'}}} \cdots \right] \nonumber\\
		  & \equiv & \frac{1}{\tau^{2}}\frac{1}{\mathcal{C}_{1}\mathcal{C}_{2}} - \frac{1}{\tau}\frac{1}{\mathcal{C}_{1}\mathcal{C}_{2}}\left[\frac{1}{\mathcal{C}_{1}} - \frac{1}{\mathcal{C}_{2}}\right]\frac{p\cdot q}{\sigma_{n+1}-\sigma_{n+2}} + \mathcal{O}(\tau^{0})
\end{eqnarray}
where we define $\mathcal{C}_{1} := \sum\limits_{a=1}^{n}\frac{p\cdot k_{a}}{\sigma_{n+1}-\sigma_{a}} $ and $\mathcal{C}_{2} := \sum\limits_{b=1}^{n}\frac{q\cdot k_{b}}{\sigma_{n+2}-\sigma_{b}}$. Product of the delta functions can be expanded as

\begin{eqnarray}\label{delta-expansion}
 \sideset{}{'}\prod\limits_{a=1}^{n} \delta \left( f_{a}\right) & = & \sideset{}{'}\prod\limits_{a=1}^{n} \delta \left(f_{a}^{n}\right) + \tau \sideset{}{'}\sum_{a=1}^{n}\left[\sideset{}{'}\prod\limits_{\substack{{b=1}\\b\ne a}}^{n} \delta \left(f_{b}^{n}\right)\right]\left(\frac{k_{a}.p}{\sigma_{a}-\sigma_{n+1}} + \frac{ k_{a}.q}{\sigma_{a}-\sigma_{n+2}} \right)\delta' \left(f_{a}^{n}\right) + \mathcal{O}(\tau^{2}) \nonumber\\
 &\equiv& \delta^{0} + \tau\delta^{1} + \mathcal{O}(\tau^{2})
\end{eqnarray}
where $f_{a}^{n} = \sum\limits_{\substack{b=1{}\\b\ne a}}^{n} \frac{k_{a}.k_{b}}{\sigma_{a}-\sigma_{b}}$. Prime denotes exclusion of any three delta functions due to $\mathbb{SL}(2,\mathbb{C})$ redundancy. Similarly we can write the integrand as a Taylor series expansion

\begin{equation}\label{I-expansion}
 I_{n+2} = I_{n+2}^{(0)} + \tau I_{n+2}^{(1)} + \cdots
\end{equation}
Therefore using Eq.(\ref{f-expansion}), Eq.(\ref{delta-expansion}) and Eq.(\ref{I-expansion}) expansion of $M_{n+2}$ in Eq.(\ref{M-expansion}) is given by

\begin{eqnarray}\label{amplitude expansion}
 M_{n+2} &\rightarrow& \frac{1}{\tau^{2}}\int \frac{\mathrm{d}^{n}\sigma}{\mathrm{vol}\mathbb{SL}(2,\mathbb{C})}\oint\limits_{|\sigma_{n+1}-\sigma_{i}|\rightarrow 0}\frac{\mathrm{d}\sigma_{n+1}}{\mathcal{C}_{1}}\oint\limits_{|\sigma_{n+2}-\sigma_{j}|\rightarrow 0}\frac{\mathrm{d}\sigma_{n+2}}{\mathcal{C}_{2}} \: \delta^{0}\: I_{n+2}^{0}(\{k,\epsilon,\sigma\}) \nonumber\\
 && + \frac{1}{\tau} \int \frac{\mathrm{d}^{n}\sigma}{\mathrm{vol}\mathbb{SL}(2,\mathbb{C})}\oint\limits_{|\sigma_{n+1}-\sigma_{i}|\rightarrow 0}\frac{\mathrm{d}\sigma_{n+1}}{\mathcal{C}_{1}}\oint\limits_{|\sigma_{n+2}-\sigma_{j}|\rightarrow 0}\frac{\mathrm{d}\sigma_{n+2}}{\mathcal{C}_{2}} \left[\delta^{(1)}I_{n+2}^{(0)} + \delta^{(0)}I_{n+2}^{(1)} \right. \nonumber\\
 && \phantom{+ \frac{1}{\tau} \int \frac{\mathrm{d}^{n}\sigma}{\mathrm{vol}\mathbb{SL}(2,\mathbb{C})}\oint_{\mathcal{C}_{1}}\frac{\mathrm{d}\sigma_{n+1}}{\mathcal{C}_{1}}\oint_{\mathcal{C}_{2}}\frac{\mathrm{d}\sigma_{n+2}}{\mathcal{C}_{2}} \left[\delta^{(1)}I_{n+2}^{(0)}\right.}\left.- \left(\frac{1}{\mathcal{C}_{1}}-\frac{1}{\mathcal{C}_{2}} \right)\frac{p\cdot q}{\sigma_{n+1}-\sigma_{n+2}}\:\delta^{0}\: I_{n+2}^{0}(\{k,\epsilon,\sigma\}) \right] \nonumber\\
 && + \: \mathcal{O}(\tau^{0})
\end{eqnarray}
It is immediately evident from the above expression that the first term readily gives product of two single soft factors.

\subsection{Degenerate Case}

The case when $|\sigma_{n+1} - \sigma_{n+2}|\sim \mathcal{O}(\tau)$ has been studied in great detail in \cite{DoubleSoftPRD, Cachazo:supp}. A new pair of variables is defined

\begin{equation}\label{sigma variables}
 \sigma_{n+1} = \rho - \frac{\xi}{2}, \qquad \sigma_{n+2} = \rho + \frac{\xi}{2}
\end{equation}
and integration of $\sigma_{n+1}$ and $\sigma_{n+2}$ now transforms as

\begin{equation}
 \mathrm{d}\sigma_{n+1}\; \mathrm{d}\sigma_{n+2}\; \delta(f_{n+1})\delta(f_{n+2})= -2\;\mathrm{d}\rho\;\mathrm{d}\xi\;\delta(f_{n+1}+f_{n+2})\delta(f_{n+1}-f_{n+2}).
\end{equation}
Actually in this situation contours of $\sigma_{n+1}$ and $\sigma_{n+2}$ approach each other at a rate $\tau$, so the contours can not be considered separately like the previous case. The soft punctures pinch together and effectively there is now a single contour of integration left. One of the delta functions is used to localize the integral to the solutions of $\xi$ and the other delta function deforms the contour of integration of $\rho$ which wraps over solutions of $\sigma_{a},\: \forall a\in \{1,2,\cdots, n\}$ . Expanding $\xi$ perturbatively in terms of $\tau$ as 

\begin{equation}
 \xi = \tau \xi_{1} + \tau^{2}\xi_{2} + \mathcal{O}(\tau^{3})
\end{equation}
we obtain an expansion of the scattering amplitude 

\begin{equation}
 M_{n+2} = -\frac{1}{\tau}\oint \frac{\mathrm{d}\rho}{2\pi i}\int\mathrm{d}\mu'_{n}\frac{\xi_{1}^{2}}{p.q \sum\limits_{b=1}^{n}\frac{k_{b}.(p+q)}{\rho-\sigma_{b}}}
	  \left(1 - \frac{\tau\xi_{1}}{2} \frac{\sum\limits_{b=1}^{n}\frac{k_{b}.(p+q)}{(\rho-\sigma_{b})^{2}}}{\sum\limits_{b=1}^{n}
	  \frac{k_{b}.(p+q)}{\rho-\sigma_{b}}} + 3\tau\frac{\xi_{2}}{\xi_{1}} + \mathcal{O}(\tau^{2}) \right)I_{n+2}.
\end{equation}
There will be an additional term which comes from taking the $\rho$ contour at infinity. 

In this case contributions come from residues evaluated at single contour integration. Hence there will be a single summation over hard legs and this term corresponds to the local processes. From the perspective of Feynman diagrams local processes occur when two soft particles are emitted either from a four point vertex or from a cubic vertex mediated via an internal propagator joined to an external leg at another cubic vertex.  

\section{Double Soft Limit in Gravity}
\label{sec:DS gravity}

In \cite{Saha} gravity amplitude in the limit when two gravitons become soft has been derived using the degenerate solution of $\xi$. There the leading term was found to be $\mathcal{O}(\tau^{-1})$. Although the result was in agreement with the Feynman diagrams at that order, we found an additional term of $\mathcal{O}(\tau^{-2})$, which was exactly product of two single soft factors, coming from one of the diagrams. It was argued that the term is absent from the CHY result because it is non-local and corresponds to the process where two soft gravitons are emitted from different hard legs. However as mentioned in Sec.(\ref{sec:CHYDS}) it will be shown in the following analysis that the non-local terms can be incorporated in the CHY result by taking into account the non-degenerate solutions. Interestingly this not only gives the leading order term at $\mathcal{O}(\tau^{-2})$ but also sub-leading order terms which are non-local as well. Moreover the final result that we obtain is consistent with the answer derived using BCFW analysis done in \cite{Klose}. 

The integrand for pure gravity theory is given in terms of reduced Pfaffian of an antisymmetric matrix in the following way

\begin{equation}\label{gravity integrand}
 I_{n} = \left(\mathrm{Pf'}\Psi_{n}(\{k,\epsilon,\sigma \})\right)^{2}.
\end{equation}
The matrix is given by

\begin{equation}
 \Psi_{n} = 
 \left(
 \begin{array}{c:c}
  A & -C^{T} \\
  \hdashline
  C^{T} & B
 \end{array}
 \right)
\end{equation}
where each of $ A, B $ and $ C $ is $ n\times n $ matrix and the components are:

\begin{equation}
\begin{aligned}[c]
A_{ab} =
\begin{cases}
\frac{k_{a}.k_{b}}{\sigma_{a} - \sigma_{b}}, & a \ne b \\
0, & a=b
\end{cases}
\end{aligned}
\qquad
\begin{aligned}[c]
B_{ab} = 
\begin{cases}
 \frac{\epsilon_{a}.\epsilon_{b}}{\sigma_{a} - \sigma_{b}}, & a \ne b \\
 0, & a=b
\end{cases}
\end{aligned}
\qquad
\begin{aligned}[c]
C_{ab} =
\begin{cases}
 \frac{\epsilon_{a}.k_{b}}{\sigma_{a}-\sigma_{b}}, & a \ne b\\
 -\sum\limits_{c \ne a}\frac{\epsilon_{a}.k_{c}}{\sigma_{a}-\sigma_{c}}, & a=b.
\end{cases}
\end{aligned}
\end{equation}
The Pfaffian of $ \Psi_{n} $ vanishes because it has a nontrivial kernel of dimension two, spanned by the vectors

\begin{equation}
 (1,1, \ldots, 1; 0,0, \ldots, 0)^{T} \quad \text{and} \quad (\sigma_{1}, \sigma_{2}, \ldots, \sigma_{n}; 0,0, \ldots, 0)^{T}.
\end{equation}
Reduced Pfaffian is defined by deleting any $i$th row and $j$th column of the above matrix with $i,\;j \in\{1,2,\cdots,n\}$,

\begin{equation}
 \mathrm{Pf'}\Psi_{n} = \frac{(-1)^{i+j}}{(\sigma_{i} - \sigma_{j})} \mathrm{Pf}(\Psi_{n})^{ij}_{ij}.
\end{equation}

We will first evaluate Eq.(\ref{amplitude expansion}). Let us label the soft momenta as $k_{n+1} = \tau p$ and $k_{n+2} = \tau q$. Then $\Psi_{n+2}$ in the gravity integrand can be expressed as

\begin{equation}\label{finite rho matrix}
 \Psi_{n+2} =
 \left(
 \begin{array}{c:c:c|c:c:c}
  (A_n)_{ab} & \frac{\tau\: k_{a}.p}{\sigma_{a} - \sigma_{n+1}} & \frac{\tau\: k_{a}.q}{\sigma_{a} - \sigma_{n+2}} & (-C_{n}^{T})_{ab} & \frac{-\epsilon_{n+1}.k_{a}}{\sigma_{n+1}-\sigma_{a}} & \frac{-\epsilon_{n+2}.k_{a}}{\sigma_{n+2}-\sigma_{a}} \\
  \hdashline\\
  \frac{\tau \: p.k_{b}}{\sigma_{n+1} - \sigma_{b}} & 0 & \frac{\tau^{2} \: p.q}{\sigma_{n+1}-\sigma_{n+2}} & \frac{-\tau \: \epsilon_{b}.p}{\sigma_{b}-\sigma_{n+1}} & -C_{n+1,n+1} & \frac{-\tau \: \epsilon_{n+2}.p}{\sigma_{n+2}-\sigma_{n+1}}\\
  \hdashline\\
  \frac{\tau \: q.k_{b}}{\sigma_{n+2} - \sigma_{b}} & \frac{\tau^{2} \: p.q}{\sigma_{n+2}-\sigma_{n+1}} & 0 & \frac{-\tau \: \epsilon_{b}.q}{\sigma_{b}-\sigma_{n+2}} & \frac{\tau \: \epsilon_{n+1}.q}{\sigma_{n+1}-\sigma_{n+2}} & -C_{n+2,n+2} \\
  \hline\\
  (C_{n})_{ab} & \frac{\tau \: \epsilon_{a}.p}{\sigma_{a}-\sigma_{n+1}} & \frac{\tau \: \epsilon_{a}.q}{\sigma_{a}-\sigma_{n+2}} & (B_{n})_{ab} & \frac{\epsilon_{a}.\epsilon_{n+1}}{\sigma_{a}-\sigma_{n+1}} & \frac{\epsilon_{a}.\epsilon_{n+2}}{\sigma_{a}-\sigma_{n+2}} \\
  \hdashline\\
  \frac{\epsilon_{n+1}.k_{b}}{\sigma_{n+1}-\sigma_{b}} & C_{n+1,n+1} & \frac{\tau \: \epsilon_{n+1}.q}{\sigma_{n+1}-\sigma_{n+2}} & \frac{\epsilon_{n+1}.\epsilon_{b}}{\sigma_{n+1}-\sigma_{b}} & 0 & \frac{\epsilon_{n+1}.\epsilon_{n+2}}{\sigma_{n+1}-\sigma_{n+2}} \\
  \hdashline\\
  \frac{\epsilon_{n+2}.k_{b}}{\sigma_{n+2}-\sigma_{b}} & \frac{\tau \: \epsilon_{n+2}.p}{\sigma_{n+2}-\sigma_{n+1}} & C_{n+2,n+2} & \frac{\epsilon_{n+2}.\epsilon_{b}}{\sigma_{n+2}-\sigma_{b}} & \frac{\epsilon_{n+2}.\epsilon_{n+1}}{\sigma_{n+2}-\sigma_{n+1}} & 0 
 \end{array}
 \right)
\end{equation}
At leading order the gravity integrand, $\text{Pf}'\left(\Psi_{n+2}\right)^{2}$ becomes

\begin{equation}
 I_{n+2}^{(0)} = \left(\sum\limits_{a}\frac{\epsilon_{n+1}\cdot k_{a}}{\sigma_{n+1}-\sigma_{a}}\right)^{2} \left(\sum\limits_{b}\frac{\epsilon_{n+2}\cdot k_{b}}{\sigma_{n+2}-\sigma_{b}}\right)^{2} I_{n}
\end{equation}
Therefore the leading order double soft factor is

\begin{eqnarray}\label{leading soft}
 S^{(0)}(p,q) & = & \frac{1}{\tau^{2}}\oint\limits_{|\sigma_{n+1}-\sigma_{i}|\rightarrow 0}\mathrm{d}\sigma_{n+1}\frac{\left(\sum\limits_{a}\frac{\epsilon_{n+1}\cdot k_{a}}{\sigma_{n+1}-\sigma_{a}}\right)^{2}}{\sum\limits_{a'}\frac{p\cdot k_{a'}}{\sigma_{n+1}-\sigma_{a'}}}\oint\limits_{|\sigma_{n+2}-\sigma_{j}|\rightarrow 0}\mathrm{d}\sigma_{n+2}\frac{\left(\sum\limits_{b}\frac{\epsilon_{n+2}\cdot k_{b}}{\sigma_{n+2}-\sigma_{b}}\right)^{2}}{\sum\limits_{b'}\frac{q\cdot k_{b'}}{\sigma_{n+2}-\sigma_{b'}}} \nonumber\\
 & = & \left(\sum\limits_{a=1}^{n}\frac{\epsilon_{n+1,\: \mu\nu}k_{a}^{\mu}k_{a}^{\nu}}{k_{a}\cdot p}\right) \times \left(\sum\limits_{b=1}^{n}\frac{\epsilon_{n+2,\: \mu\nu}k_{b}^{\mu}k_{b}^{\nu}}{k_{a}\cdot q}\right)\nonumber\\
 & = & S^{(0)}(p) S^{(0)}(q)
\end{eqnarray}
which is product of two leading order single soft factors \cite{Weinberg} as expected. This term is non-local and satisfies the properties of gauge invariance. It is obvious that a generic result holds for leading factor in multiple soft emissions which is given by product of that many single soft factors\cite{VolovichZlotnikov}.  

Next we consider the sub-leading terms in Eq.(\ref{amplitude expansion}). Let us look at the term with $\sigma_{n+1} - \sigma_{n+2}$ in the denominator. We get an expression 

\begin{equation}
 \oint\limits_{|\sigma_{n+1}-\sigma_{i}|\rightarrow 0}\mathrm{d}\sigma_{n+1}\frac{\left(\sum\limits_{a=1}^{n}\frac{\epsilon_{n+1}\cdot k_{a}}{\sigma_{n+1}-\sigma_{a}}\right)^{2}}{\sum\limits_{a'=1}^{n}\frac{p\cdot k_{a'}}{\sigma_{n+1}-\sigma_{a'}}}\oint\limits_{|\sigma_{n+2}-\sigma_{j}|\rightarrow 0}\mathrm{d}\sigma_{n+2}\frac{\left(\sum\limits_{b=1}^{n}\frac{\epsilon_{n+2}\cdot k_{b}}{\sigma_{n+2}-\sigma_{b}}\right)^{2}}{\left(\sum\limits_{b'=1}^{n}\frac{q\cdot k_{b'}}{\sigma_{n+2}-\sigma_{b'}}\right)^{2}} \frac{p\cdot q}{\sigma_{n+1}-\sigma_{n+2}}.
\end{equation}
If we do the $\sigma_{n+2}$ integration first we will find that the contour integral does not contain any pole and therefore there is no residue. Also since $ |\sigma_{n+1} - \sigma_{n+2}|\nrightarrow 0$, so $\sigma_{n+1}$ is outside the contour of $\sigma_{n+2}$ which wraps over the solutions of $\sigma_{a}, \: \forall a\in \{1,2,\cdots, n\}$. Hence the contour integration vanishes. Thus we see the last term in Eq.(\ref{amplitude expansion}) at $\mathcal{O}(\tau^{-1})$ drops out. The non vanishing contributions come from the remaining terms which are combinations $\delta^{(1)}I_{n+2}^{(0)}$ and $\delta^{(0)}I_{n+2}^{(1)}$ as will be explained below.

Sub-leading soft factor for gravity is given by \cite{CachazoStrominger} 

\begin{equation}
 S^{(1)}(p) = \sum\limits_{a=1}^{n}\frac{\epsilon_{n+1, \mu\nu}k_{a}^{\mu}p_{\rho}\hat{J}_{a}^{\rho,\nu}}{p\cdot k_{a}}, \qquad
 S^{(1)}(q) = \sum\limits_{a=1}^{n}\frac{\epsilon_{n+2, \mu\nu}k_{a}^{\mu}q_{\rho}\hat{J}_{a}^{\rho,\nu}}{q\cdot k_{a}}
\end{equation}
where $\hat{J}$ is a first order differential operator which acts on both momenta and polarizations. In the subsequent steps we will closely follow the analysis of \cite{Afkhami}. Acting $ S^{(1)}(p)$ on $M_{n}$ we get

\begin{eqnarray}\label{sub-leading terms}
 S^{(1)}(p) M_{n} & = & \int \frac{\mathrm{d}^{n}\sigma}{\mathrm{vol}\mathbb{SL}(2,\mathbb{C})} \sideset{}{'}\sum\limits_{l}\left[\sideset{}{'}\prod\limits_{\substack{a\\a\ne l}} \delta \left(f_{a}^{n}\right)\right]\delta'\left(f_{l}^{n}\right)\sum\limits_{\substack{b=1\\b\ne l}}^{n}\frac{1}{\sigma_{l}-\sigma_{b}}\left[2\epsilon_{n+1}\cdot k_{b} \; \epsilon_{n+1}\cdot k_{l} \right. \nonumber\\
 && \left.\phantom{\int \frac{\mathrm{d}^{n}\sigma}{\mathrm{vol}\mathbb{SL}(2,\mathbb{C})} \sideset{}{'}\sum\limits_{l}\left[\sideset{}{'}\prod\limits_{\substack{a\\a\ne l}} \delta \left(f_{a}^{n}\right)\right]\delta'\left(f_{l}^{n}\right)}- \frac{\left(\epsilon_{n+1}\cdot k_{b}\right)^{2} p\cdot k_{l}}{p\cdot k_{b}} - \frac{\left(\epsilon_{n+1}\cdot k_{l}\right)^{2}p\cdot k_{b}}{p\cdot k_{l}}\right] I_{n} \nonumber\\
 && + \int \frac{\mathrm{d}^{n}\sigma}{\mathrm{vol}\mathbb{SL}(2,\mathbb{C})}\sideset{}{'}\prod\limits_{a}\delta \left(f_{a}^{n}\right)S^{(1)}(p)I_{n}
\end{eqnarray}
Let us now focus on the $\delta^{(1)}I_{n+2}^{(0)}$ term of Eq.(\ref{amplitude expansion}) and compare with the first term of Eq.(\ref{sub-leading terms}). 

\begin{eqnarray}\label{S1-delta}
&& \int \frac{\mathrm{d}^{n}\sigma}{\mathrm{vol}\mathbb{SL}(2,\mathbb{C})}\oint\limits_{|\sigma_{n+1}-\sigma_{l}|\rightarrow 0}\frac{\mathrm{d}\sigma_{n+1}}{\sum\limits_{a=1}^{n} \frac{\tau k_{a}.p}{\sigma_{n+1}-\sigma_{a}}}\oint\limits_{|\sigma_{n+2}-\sigma_{m}|\rightarrow 0}\frac{\mathrm{d}\sigma_{n+2}}{\sum\limits_{b=1}^{n} \frac{\tau k_{b}.q}{\sigma_{n+2}-\sigma_{b}}} \delta^{(1)}I_{n+2}^{(0)} \nonumber\\
& = &\int \frac{\mathrm{d}^{n}\sigma}{\mathrm{vol}\mathbb{SL}(2,\mathbb{C})} \oint\limits_{|\sigma_{n+1}-\sigma_{l}|\rightarrow 0}\frac{\mathrm{d}\sigma_{n+1}}{\sum\limits_{i=1}^{n} \frac{\tau k_{i}.p}{\sigma_{n+1}-\sigma_{i}}}\oint\limits_{|\sigma_{n+2}-\sigma_{m}|\rightarrow 0}\frac{\mathrm{d}\sigma_{n+2}}{\sum\limits_{j=1}^{n} \frac{\tau k_{j}.q}{\sigma_{n+2}-\sigma_{j}}}\left(\sum\limits_{a'}\frac{\epsilon_{n+1}\cdot k_{a'}}{\sigma_{n+1}-\sigma_{a'}}\right)^{2} \left(\sum\limits_{b'}\frac{\epsilon_{n+2}\cdot k_{b'}}{\sigma_{n+2}-\sigma_{b'}}\right)^{2} \nonumber\\
&& \phantom{\oint\limits_{|\sigma_{n+1}-\sigma_{l}|\rightarrow 0}\frac{\mathrm{d}\sigma_{n+1}}{\sum\limits_{i=1}^{n} \frac{\tau k_{i}.p}{\sigma_{n+1}-\sigma_{i}}}} \times \tau \sideset{}{'}\sum_{a}\left[\sideset{}{'}\prod\limits_{\substack{{b}\\b\ne a}} \delta \left(f_{b}^{n}\right)\right]\left(\frac{k_{a}.p}{\sigma_{a}-\sigma_{n+1}} + \frac{ k_{a}.q}{\sigma_{a}-\sigma_{n+2}} \right)\delta' \left(f_{a}^{n}\right)I_{n}
\end{eqnarray}
Now using

\begin{eqnarray}
 &&\oint\limits_{|\sigma_{n+1}-\sigma_{l}|\rightarrow 0}\mathrm{d}\sigma_{n+1}\frac{k_{a}.p}{\sigma_{a}-\sigma_{n+1}}\frac{\left(\sum\limits_{a'}\frac{\epsilon_{n+1}\cdot k_{a'}}{\sigma_{n+1}-\sigma_{a'}}\right)^{2}}{{\sum\limits_{i=1}^{n} \frac{\tau k_{i}.p}{\sigma_{n+1}-\sigma_{i}}}} \nonumber\\
 &=& \frac{1}{\tau}\sum\limits_{\substack{b=1\\b\ne a}}^{n}\left[-\frac{k_{a}\cdot p}{\sigma_{a}-\sigma_{b}}\frac{\left(\epsilon_{n+1}\cdot k_{b}\right)^{2}}{k_{b}\cdot p} +2\frac{\epsilon_{n+1}\cdot k_{a}\; \epsilon_{n+1}\cdot k_{b}}{\sigma_{a}-\sigma_{b}} - \frac{\left(\epsilon_{n+1}\cdot k_{a}\right)^{2}k_{b}\cdot p}{\left(\sigma_{a}-\sigma_{b}\right)k_{a}\cdot p}\right]
\end{eqnarray}
and comparing with Eq.(\ref{sub-leading terms}) it is evident that Eq.(\ref{S1-delta}) becomes

\begin{equation}\label{delta1}
 S^{(0)}(q)\int \frac{\mathrm{d}^{n}\sigma}{\mathrm{vol}\mathbb{SL}(2,\mathbb{C})}\left( S^{(1)}(p)\left[\sideset{}{'}\prod\limits_a \delta \left(f_{a}^{n}\right)\right]\right)I_{n} + (p\leftrightarrow q).
\end{equation}
In the rest of the analysis to make our calculations easier we will choose the following gauge fixing conditions  

\begin{eqnarray}\label{gauge}
  \epsilon_{n+1}\cdot q & = & 0 \nonumber\\
  \epsilon_{n+2}\cdot p &= & 0 \nonumber\\
  \epsilon_{a}\cdot q & = & 0,  \qquad \forall a \in \{1,2,\cdots,n\}.
\end{eqnarray}
It is to be noted that our final results remain unaffected by this choice of gauge conditions. These conditions allow us to reduce the number of terms appearing in the intermediate steps of our calculations, nevertheless one can also do similar analysis without fixing any gauge condition as is done in \cite{Kalousios}. Now our task is to find $I_{n+2}^{(1)}$ and calculate the remaining term in Eq.(\ref{amplitude expansion}). Taking derivative of the determinant and using Eq.(\ref{gauge}) we get

\begin{eqnarray}\label{expansion1}
 \frac{\partial I_{n+2}}{\partial\tau}|_{\tau=0}& = & \sum\limits_{a=1}^{n}\left[(-1)^{n+a+1}\frac{k_{a}\cdot p}{\sigma_{a}-\sigma_{n+1}}\tilde{\Psi}_{n+1}^{a} + (-1)^{n+a} \frac{k_{a}\cdot q}{\sigma_{a}-\sigma_{n+2}}\tilde{\Psi}_{n+2}^{a} + (-1)^{n}\frac{\epsilon_{a}\cdot p}{\sigma_{a}-\sigma_{n+1}}\tilde{\Psi}_{n+a}^{a} \right. \nonumber\\
 && \phantom{\sum\limits_{a=1}^{n}\left[\right.}\left. + (-1)^{n+1}\frac{\epsilon_{a}\cdot p}{\sigma_{a}-\sigma_{n+1}}\tilde{\Psi}_{a}^{n+2+a} + (-1)^{a+1}\frac{\epsilon_{a}\cdot p}{\sigma_{a}-\sigma_{n+1}}\tilde{\Psi}_{n+1}^{n+2+a} \right]\nonumber\\
 && + (-1)^{n}\left[C_{n+1,n+1}\tilde{\Psi}^{2n+3}_{n+1} + C_{n+2,n+2}\tilde{\Psi}^{2n+4}_{n+2}\right]
\end{eqnarray}
where $\tilde{\Psi}^{a}_{b}$ denotes determinant of the reduced matrix with $a$th row and $b$th column removed. After expanding the reduced determinants the above equation can be written as

\begin{eqnarray}\label{subleading determinant}
 I_{n+2}^{(1)}& = & \left(C_{n+2,n+2}\right)^{2}C_{n+1,n+1}\sum\limits_{a=1}^{n}\sum\limits_{b=1}^{n}\left[\frac{k_{a}\cdot p}{\sigma_{a}-\sigma_{n+1}}\left((-1)^{a+b+1}\frac{\epsilon_{n+1}\cdot k_{b}}{\sigma_{n+1}-\sigma_{b}}\Psi^{a}_{b} +(-1)^{n+a+b+1}\frac{\epsilon_{n+1}\cdot\epsilon_{b}}{\sigma_{n+1}-\sigma_{b}}\Psi^{a}_{n+b} \right) \right.\nonumber\\
 && \phantom{\left(C_{n+2,n+2}\right)^{2}C_{n+1,n+1}\sum\limits_{a=1}^{n}\sum\limits_{b=1}^{n}\left[\right.}\left. + \frac{p\cdot k_{b}}{\sigma_{n+1}-\sigma_{b}}\left((-1)^{a+b+1}\frac{\epsilon_{n+1}\cdot k_{a}}{\sigma_{n+1}-\sigma_{a}}\Psi^{a}_{b} + (-1)^{n+a+b}\frac{\epsilon_{a}\cdot\epsilon_{n+1}}{\sigma_{a}-\sigma_{n+1}}\Psi^{n+a}_{b} \right) \right. \nonumber\\
 && \phantom{\left(C_{n+2,n+2}\right)^{2}C_{n+1,n+1}\sum\limits_{a=1}^{n}\sum\limits_{b=1}^{n}\left[\right.}\left. +  \frac{\epsilon_{b}\cdot p}{\sigma_{b}-\sigma_{n+1}}\left((-1)^{n+a+b}\frac{\epsilon_{n+1}\cdot k_{a}}{\sigma_{n+1}-\sigma_{a}}\Psi^{a}_{n+b} + (-1)^{a+b+1}\frac{\epsilon_{a}\cdot\epsilon_{n+1}}{\sigma_{a}-\sigma_{n+1}}\Psi^{n+a}_{n+b} \right)\right.\nonumber\\
 && \phantom{\left(C_{n+2,n+2}\right)^{2}C_{n+1,n+1}\sum\limits_{a=1}^{n}\sum\limits_{b=1}^{n}\left[\right.} + \left. \frac{\epsilon_{a}\cdot p}{\sigma_{a}-\sigma_{n+1}}\left((-1)^{n+a+b+1}\frac{\epsilon_{n+1}\cdot k_{b}}{\sigma_{n+1}-\sigma_{b}}\Psi^{n+a}_{b} + (-1)^{a+b+1}\frac{\epsilon_{n+1}\cdot\epsilon_{b}}{\sigma_{n+1}-\sigma_{b}}\Psi^{n+a}_{n+b}\right)\right]\nonumber\\
 && + (-1)^{n}\left(C_{n+1,n+1}\right)^{2}\left(C_{n+2,n+2}\right)^{2}\sum\limits_{a=1}^{n}\frac{\epsilon_{a}\cdot p}{\sigma_{a}-\sigma_{n+1}}\left(\Psi_{n+a}^{a} - \Psi_{a}^{n+a}\right)\nonumber\\
 && + \left(C_{n+1,n+1}\right)^{2}C_{n+2,n+2}\sum\limits_{a=1}^{n}\sum\limits_{b=1}^{n} \left[\frac{k_{a}\cdot q}{\sigma_{a}-\sigma_{n+2}}\left((-1)^{a+b+1}\frac{\epsilon_{n+2}\cdot k_{b}}{\sigma_{n+2}-\sigma_{b}}\Psi^{a}_{b} +(-1)^{n+a+b+1}\frac{\epsilon_{n+2}\cdot\epsilon_{b}}{\sigma_{n+2}-\sigma_{b}}\Psi^{a}_{n+b} \right) \right.\nonumber\\
 &&\phantom{\left(C_{n+1,n+1}\right)^{2}C_{n+2,n+2}\sum\limits_{a=1}^{n}\sum\limits_{b=1}^{n}\left[\right.}\left. + \frac{q\cdot k_{b}}{\sigma_{n+2}-\sigma_{b}}\left((-1)^{a+b+1}\frac{\epsilon_{n+2}\cdot k_{a}}{\sigma_{n+2}-\sigma_{a}}\Psi^{a}_{b} + (-1)^{n+a+b}\frac{\epsilon_{a}\cdot\epsilon_{n+2}}{\sigma_{a}-\sigma_{n+2}}\Psi^{n+a}_{b} \right)\right].
\end{eqnarray}
Substituting $I_{n+2}^{(1)}$ into the relevant term in Eq.(\ref{amplitude expansion}) and doing the contour integrals over $\sigma_{n+1}$  and $\sigma_{n+2}$ we get

\begin{eqnarray}\label{S1-integrand}
 S^{(0)}(q)\sum\limits_{a=1}^{n}\sum\limits_{\substack{b=1\\b\ne a}}^{n}\frac{2}{\sigma_{a}-\sigma_{b}}&\times &\left[\left(\frac{\epsilon_{n+1}\cdot k_{a}}{p\cdot k_{a}}-\frac{\epsilon_{n+1}\cdot k_{b}}{p\cdot k_{b}} \right)\left(p\cdot k_{a}\right)\left(\epsilon_{n+1}\cdot k_{b}\right)(-1)^{a+b}\Psi^{a}_{b} \right. \nonumber\\
 & - & \left. \left(\frac{\epsilon_{n+1}\cdot k_{a}}{p\cdot k_{a}}-\frac{\epsilon_{n+1}\cdot k_{b}}{p\cdot k_{b}}\right)\left(\epsilon_{n+1}\cdot\epsilon_{a}\right)\left(\epsilon_{b}\cdot p\right)(-1)^{a+b}\Psi_{n+b}^{n+a} \right. \nonumber\\
  & +&  \left. \left(\frac{\epsilon_{n+1}\cdot k_{a}}{p\cdot k_{a}}-\frac{\epsilon_{n+1}\cdot k_{b}}{p\cdot k_{b}} \right)\bigl\{\left(p\cdot k_{a}\right)\left(\epsilon_{n+1}\cdot \epsilon_{b}\right)- \left(\epsilon_{b}\cdot p\right)\left(\epsilon_{n+1}\cdot k_{a}\right)\bigr\}(-1)^{n+a+b}\Psi^{a}_{n+b} \right.\nonumber\\
  & +& \left.  \biggl\{\left(\frac{\epsilon_{n+1}\cdot k_{a}}{p\cdot k_{a}}-\frac{\epsilon_{n+1}\cdot k_{b}}{p\cdot k_{b}} \right) \left(\epsilon_{a}\cdot p\right)\left(\epsilon_{n+1}\cdot k_{b}\right) \right.\nonumber\\
  && \hspace{10mm} \left.  + \left(\frac{\epsilon_{n+1}\cdot k_{b}}{p\cdot k_{a}}-\frac{\left(p\cdot k_{b}\right)\left(\epsilon_{n+1}\cdot k_{a}\right)}{\left(p\cdot k_{a}\right)^{2}} \right)\left(p\cdot k_{a}\right)\left(\epsilon_{n+1}\cdot \epsilon_{a}\right)\biggr\}(-1)^{n}\Psi^{a}_{n+a}\right]\nonumber\\
  +\; S^{(0)}(p)\sum\limits_{a=1}^{n}\sum\limits_{\substack{b=1\\b\ne a}}^{n}\frac{2}{\sigma_{a}-\sigma_{b}}&\times &\left[\left(\frac{\epsilon_{n+2}\cdot k_{a}}{q\cdot k_{a}}-\frac{\epsilon_{n+2}\cdot k_{b}}{q\cdot k_{b}} \right)\left(q\cdot k_{a}\right)\left(\epsilon_{n+2}\cdot k_{b}\right)(-1)^{a+b}\Psi^{a}_{b} \right. \nonumber\\
  & +&  \left. \left(\frac{\epsilon_{n+2}\cdot k_{a}}{q\cdot k_{a}}-\frac{\epsilon_{n+2}\cdot k_{b}}{q\cdot k_{b}} \right)\left(q\cdot k_{a}\right)\left(\epsilon_{n+2}\cdot \epsilon_{b}\right)(-1)^{n+a+b}\Psi^{a}_{n+b} \right.\nonumber\\
  & +& \left.  \left(\frac{\epsilon_{n+2}\cdot k_{b}}{q\cdot k_{a}}-\frac{\left(q\cdot k_{b}\right)\left(\epsilon_{n+2}\cdot k_{a}\right)}{\left(q\cdot k_{a}\right)^{2}} \right)\left(q\cdot k_{a}\right)\left(\epsilon_{n+2}\cdot \epsilon_{a}\right)(-1)^{n}\Psi^{a}_{n+a}\right].
\end{eqnarray}
Details of the above calculations are provided in Sec.(\ref{App: calculations}). Finally it can be shown that Eq.(\ref{S1-integrand}) is equal to 

\begin{equation}\label{integrand1}
 S^{(0)}(q)\int \frac{\mathrm{d}^{n}\sigma}{\mathrm{vol}\mathbb{SL}(2,\mathbb{C})}\sideset{}{'}\prod\limits_a \delta \left(f_{a}^{n}\right)\left[S^{(1)}(p)I_{n}\right] + (p\leftrightarrow q)
\end{equation}
Adding together Eq.(\ref{delta1}) and Eq.(\ref{integrand1}) the non-contact subleading terms of the double soft factor are obtained

\begin{equation}\label{subleading soft}
 S^{(0)}(p)S^{(1)}(q) + S^{(0)}(q)S^{(1)}(p).
\end{equation}

This completes our analysis of nondegenerate solutions leading to nonlocal terms. Since $S^{(0)}$ and $S^{(1)}$ are individually gauge invariant, each of the terms also remains gauge invariant. Also leading and subleading single soft factors do not depend on the helicity of soft graviton and hence expressions (\ref{leading soft}) and (\ref{subleading soft}) are also independent of helicities of the soft gravitons.

The contribution from the degenerate solution has been derived in \cite{Saha}. It gives a contact term at subleading order

\begin{eqnarray}\label{contact}
 S^{(1)}(p,q) &=& -\frac{1}{\tau}\sum\limits_{a=1}^{n}\left[ \frac{1}{k_{a}.(p+q) \: p.q}\biggl\{ -(\epsilon_{n+1}.\epsilon_{n+2})^{2}\: k_{a}.p\: k_{a}.q 
	      +2 \: \epsilon_{n+1}.\epsilon_{n+2} \left(\epsilon_{n+1}.q \:\epsilon_{n+2}.k_{a} \: k_{a}.p + \epsilon_{n+1}.k_{a} \: \epsilon_{n+2}.p \: k_{a}.q \right) \right. \nonumber \\
	      && \left. \phantom{-\frac{1}{\tau}\sum\limits_{a=1}^{n}\left[\frac{1}{k_{a}.(p+q) \: p.q}\biggl\{\right.} -2 \: \epsilon_{n+1}.q \: \epsilon_{n+2}.p \: \epsilon_{n+1}.k_{a} \: \epsilon_{n+2}.k_{a} 
	      + (\epsilon_{n+1}.q)^{2} \: (\epsilon_{n+2}.k_{a})^{2} + (\epsilon_{n+1}.k_{a})^{2}\:(\epsilon_{n+2}.p)^{2} \biggr\} \right. \nonumber \\
	      && \left. \phantom{-\frac{1}{\tau}\sum\limits_{a=1}^{n}\left[ \right.} - \frac{1}{p.q} \biggl\{ \frac{(\epsilon_{n+1}.q)^{2}\:(\epsilon_{n+2}.k_{a})^{2}}{k_{a}.q} 
	      + \frac{(\epsilon_{n+1}.k_{a})^{2}\: (\epsilon_{n+2}.p)^{2}}{k_{a}.p} \biggr\} \right. \nonumber\\
	      && \left. \phantom{-\frac{1}{\tau}\sum\limits_{a=1}^{n}\left[ \right.} + \frac{1}{k_{a}.(p+q)} \biggl\{ -2 \: \epsilon_{n+1}.\epsilon_{n+2} \: \epsilon_{n+1}.k_{a} \:\epsilon_{n+2}.k_{a}
	      + 2 \: \epsilon_{n+1}.k_{a}\: \epsilon_{n+2}.k_{a} \left( \frac{\epsilon_{n+1}.q \: \epsilon_{n+2}.k_{a}}{k_{a}.q} \:+ \right. \right.\nonumber\\
	      &&\left.\left. \hspace{70mm}  \frac{\epsilon_{n+1}.k_{a} \: \epsilon_{n+2}.p}{k_{a}.p}\right)  - \frac{(\epsilon_{n+1}.k_{a})^{2} \: (\epsilon_{n+2}.k_{a})^{2} \: p.q}{k_{a}.p \: k_{a}.q} \biggr\} \right].
\end{eqnarray}
Interestingly this contact term vanishes when both the soft particles are of same helicity and survives when they are different. It can be shown in the spinor helicity notations if we choose $\epsilon_{n+1,\: \alpha\dot{\alpha}}^{(+)} = \frac{\lambda_{q,\:\alpha}\tilde{\lambda}_{p,\: \dot{\alpha}}}{\langle qp\rangle}$ and $\epsilon_{n+2,\: \alpha\dot{\alpha}}^{(-)} = \frac{\lambda_{q,\:\alpha}\tilde{\lambda}_{p,\: \dot{\alpha}}}{[pq]}$, then the above expression reduces to that of \cite{Klose}

\begin{equation}
 \text{DSL}^{(1)}(p^{+},\: q^{-})\mid_{\text{c}} = \frac{1}{2\:p\cdot q}\sum\limits_{a=1}^{n} \frac{[p\;a]^{3}\langle q\;a\rangle^{3}}{[q\;a]\langle p\; a\rangle}\frac{1}{2\; k_{a}\cdot (p+q)}.
\end{equation}
Therefore we see double soft factors of gravity amplitude upto subleading order are given by

\begin{eqnarray}\label{DSfactor}
 S^{(0)}(p,q) & = & S^{(0)}(p)S^{(0)}(q) \nonumber\\
 S^{(1)}(p,q) & = & S^{(0)}(p)S^{(1)}(q) + S^{(0)}(q)S^{(1)}(p) + S^{(1)}(p,q)\mid_{\text{contact}}.
\end{eqnarray}

\section{Discussions}
\label{sec:discussion}

In \cite{DoubleSoftPRD, Cachazo:supp} double soft behavior of large class of theories containing scalars have been explored in great detail. It was shown that in sGal, DBI, EMS, NLSM and YMS contributions from nondegenerate solutions are suppressed compared to degenerate ones. In this analysis we find in case of gravity opposite thing happens: leading order double soft factor comes from nondegenerate solutions whereas at subleading order both degenerate as well as nondegenerate solutions contribute. 

Moreover this analysis improves our understanding of local and nonlocal terms in terms of punctures on the complex spheres. 

\begin{itemize}
 \item When the soft punctures coalesce together we get unique degenerate solution such that effectively we have a single contour integration and this gives us the contact term. This term corresponds to the scattering processes where soft gravitons are emitted from the same external hard legs. As was explained in \cite{Saha} the local term contains a four point vertex and combinations of three point vertices.
 
 \item The noncontact terms appear as a result of the nondegenerate solutions where there are two separate and independent contour integrals, performing each of these integrals we obtain single soft factors and this explains why these terms always occur as product of soft factors. Noncontact  terms relate to the scattering processes where soft gravitons can be emitted from different hard legs.
\end{itemize}

Another important property of the derived soft factors (\ref{DSfactor}) is the independence of spacetime dimensions. 

\section{Conclusion}
\label{sec:conclusion}

In this paper we derived double soft factors of pure gravity amplitude at leading and subleading orders from CHY formalism. It is now clear that both degenerate and nondegenerate solutions have contributions at subleading order while leading order soft factor comes from taking into account the nondegenerate solutions only. Our finding adds to the result of \cite{Saha}, which was incomplete because noncontact terms were left out. Here we presented detail explanations for noncontact and contact soft factors in terms of nondegenerate and degenerate solutions. 

Leading and subleading factors (\ref{DSfactor}) are valid at tree level and hold in any arbitrary dimensions of spacetime. Loop corrections to single soft factors have been studied in \cite{Bern:loopcor, He:loopcor, Bianchi:loops}. It can be shown that leading soft factor is protected from loop corrections whereas subleading ones receive corrections at loop level. It will be interesting future work to see how the double soft factors behave under loop corrections. 

Soft theorems of gravity and Yang-Mills are manifestations of asymptotic symmetries of space time. In \cite{Strominger:ST, Strominger:Scat, Strominger:2014pwa, Campiglia:2015yka, Campiglia:2014yka, Campiglia:2016jdj, Campiglia:2015kxa} relations between Weinberg's soft theorem and BMS symmetries at asymptotic null infinity have been established. It will be interesting to study the asymptotic symmetries, if present, for double soft gravity theorem. Current-current algebra for double soft gluon amplitude at null infinity has been studied in \cite{He:2015zea, McLoughlin:2016uwa} where it was shown to produce a level zero Kac-Moody current. Similar such studies may be done for double soft graviton amplitude also. The noncontact terms in the gravity amplitude may be due to the action of two supertranslation operators at leading order and supertranslation and superrotaion operators together at subleading order. However the contact term makes situation more interesting and some nontrivial symmetry may be responsible for its presence\footnote{We thank Alok Laddha for providing his valuable insights regarding this issue.}.   

\section{Acknowledgements}

I am indebted to Alok Laddha for lots of valuable discussions, for providing clarifications regarding some issues with gauge choices and for helping me with this work. I would like to thank him for carefully going through several versions of the manuscript and pointing out errors. I am grateful to S. Kalyana Rama for his valuable comments and suggestions on the draft. I would also like to thank Song He for helpful discussions during Asian Winter School on Strings, Particles and Cosmology, 2017, held in China. I have also been immensely benefited from the innumerable discussions with Madhusudhan Raman and Prashanth Raman. I would also like to thank the anonymous referee for providing valuable comments and suggestions.

\appendix
\section{Details of calculations}
\label{App: calculations}

The reduced determinants in Eq.(\ref{expansion1}) are given by

\begin{eqnarray}
 \tilde{\Psi}^{a}_{n+1} & = & (-1)^{n}\left(C_{n+2,n+2}\right)^{2}C_{n+1,n+1}\sum\limits_{b=1}^{n}\left[(-1)^{b}\frac{\epsilon_{n+1}\cdot k_{b}}{\sigma_{n+1}-\sigma_{b}} \Psi^{a}_{b} + (-1)^{n+b}\frac{\epsilon_{n+1}\cdot\epsilon_{b}}{\sigma_{n+1}-\sigma_{b}}\Psi^{a}_{n+b} \right] \nonumber\\
 \tilde{\Psi}^{a}_{n+2} & = & (-1)^{n+1}\left(C_{n+1,n+1}\right)^{2}C_{n+2,n+2}\sum\limits_{b=1}^{n}\left[(-1)^{b}\frac{\epsilon_{n+2}\cdot k_{b}}{\sigma_{n+2}-\sigma_{b}} \Psi^{a}_{b} + (-1)^{n+b}\frac{\epsilon_{n+2}\cdot\epsilon_{b}}{\sigma_{n+2}-\sigma_{b}}\Psi^{a}_{n+b} \right] \nonumber\\
 \tilde{\Psi}^{2n+3}_{n+1} & = & \left(C_{n+2,n+2}\right)^{2}\sum\limits_{a=1}^{n}\sum\limits_{b=1}^{n}\left[\frac{p\cdot k_{b}}{\sigma_{n+1}-\sigma_{b}}\left((-1)^{n+a+b+1}\frac{\epsilon_{n+1}\cdot k_{a}}{\sigma_{n+1}-\sigma_{a}}\Psi^{a}_{b} + (-1)^{a+b}\frac{\epsilon_{a}\cdot\epsilon_{n+1}}{\sigma_{a}-\sigma_{n+1}}\Psi^{n+a}_{b} \right) \right. \nonumber\\
 && \phantom{\left(C_{n+2,n+2}\right)^{2}\sum\limits_{a=1}^{n}\sum\limits_{b=1}^{n}\left[\right.} \left. + \frac{\epsilon_{b}\cdot p}{\sigma_{b}-\sigma_{n+1}}\left( (-1)^{a+b}\frac{\epsilon_{n+1}\cdot k_{a}}{\sigma_{n+1}-\sigma_{a}}\Psi^{a}_{n+b} + (-1)^{n+a+b+1}\frac{\epsilon_{a}\cdot\epsilon_{n+1}}{\sigma_{a}-\sigma_{n+1}}\Psi^{n+a}_{n+b} \right) \right] \nonumber\\
 \tilde{\Psi}^{2n+4}_{n+2} & = & \left(C_{n+1,n+1}\right)^{2}\sum\limits_{a=1}^{n}\sum\limits_{b=1}^{n}\frac{q\cdot k_{b}}{\sigma_{n+2}-\sigma_{b}}\left[(-1)^{n+a+b+1}\frac{\epsilon_{n+2}\cdot k_{a}}{\sigma_{n+2}-\sigma_{a}}\Psi^{a}_{b} + (-1)^{a+b}\frac{\epsilon_{a}\cdot\epsilon_{n+2}}{\sigma_{a}-\sigma_{n+2}}\Psi^{n+a}_{b} \right] \nonumber\\
 \tilde{\Psi}_{n+1}^{n+2+a} & = & (-1)^{n}C_{n+1,n+1}\left(C_{n+2,n+2}\right)^{2}\sum\limits_{b=1}^{n}\left[(-1)^{b}\frac{\epsilon_{n+1}\cdot k_{b}}{\sigma_{n+1}-\sigma_{b}}\Psi_{b}^{n+a} + (-1)^{n+b}\frac{\epsilon_{n+1}\cdot\epsilon_{b}}{\sigma_{n+1}-\sigma_{b}}\Psi_{n+b}^{n+a}\right]\nonumber\\
 \tilde{\Psi}_{n+a}^{a} & = & \left(C_{n+1,n+1}\right)^{2}\left(C_{n+2,n+2}\right)^{2}\Psi_{n+a}^{a} \nonumber\\
 \tilde{\Psi}_{a}^{n+2+a} & = & \left(C_{n+1,n+1}\right)^{2}\left(C_{n+2,n+2}\right)^{2}\Psi_{a}^{n+a} 
\end{eqnarray}
Plugging back these expressions we obtain Eq.(\ref{subleading determinant}).Now using the relation $\Psi^{a}_{b} = - \Psi^{b}_{a}$ which holds because of the antisymmetric property and the following result

\begin{eqnarray}\label{I-integral}
 I_{ab} & = & \oint\mathrm{d}\sigma_{n+1}\frac{\sum\limits_{c=1}^{n}\frac{\epsilon_{n+1}\cdot k_{c}}{\sigma_{n+1}-\sigma_{c}}}{\sum\limits_{d=1}^{n}\frac{p\cdot k_{d}}{\sigma_{n+1}-\sigma_{d}}}\frac{1}{\left(\sigma_{n+1}-\sigma_{a}\right)\left(\sigma_{n+1}-\sigma_{b}\right)}\nonumber\\
 & = & 
        \begin{cases}
         \frac{\epsilon_{n+1}\cdot k_{a}}{p\cdot k_{a}}\frac{1}{\sigma_{a}-\sigma_{b}} + \frac{\epsilon_{n+1}\cdot k_{b}}{p\cdot k_{b}}\frac{1}{\sigma_{b}-\sigma_{a}},  \qquad & a\ne b \\
         \sum\limits_{\substack{d=1\\d\ne a}}^{n}\frac{1}{\sigma_{a}-\sigma_{d}}\left[\frac{\epsilon_{n+1}\cdot k_{d}}{p\cdot k_{a}}-\frac{\left(p\cdot k_{d}\right)\left(\epsilon_{n+1}\cdot k_{a}\right)}{\left(p\cdot k_{a}\right)^{2}} \right], \qquad & a=b
        \end{cases}
\end{eqnarray}
we get the first part in Eq.(\ref{S1-integrand}) multiplying $S^{(0)}(q)$ factor. Similarly one can obtain the other part.

Now we consider the action of $S^{(1)}$ factor on the determinant $I_{n}$.

\begin{eqnarray}\label{S1-In}
 S^{(1)}(p)I_{n} &= & 2\sum\limits_{\substack{a=1\\a\ne b}}^{n}\sum\limits_{b=1}^{n}\frac{1}{\sigma_{a}-\sigma_{b}}\left(S^{(1)}_{b}\left(k_{b}\cdot k_{a}\right)(-1)^{a+b}\Psi^{a}_{b} + S^{(1)}_{b}\left(\epsilon_{b}\cdot\epsilon_{a}\right)(-1)^{a+b}\Psi^{n+a}_{n+b} \right. \nonumber\\
 && \left. \phantom{2\sum\limits_{\substack{a=1\\a\ne b}}^{n}\sum\limits_{b=1}^{n}\frac{1}{\sigma_{a}-\sigma_{b}}\left(\right.} + \left[S^{(1)}_{b}\left(\epsilon_{b}\cdot k_{a}\right) + S^{(1)}_{a}\left(k_{a}\cdot\epsilon_{b}\right)\right](-1)^{n+a+b}\Psi^{a}_{n+b} \right. \nonumber\\
 &&\left. \phantom{2\sum\limits_{\substack{a=1\\a\ne b}}^{n}\sum\limits_{b=1}^{n}\frac{1}{\sigma_{a}-\sigma_{b}}\left(\right.} + \left[S^{(1)}_{b}\left(k_{b}\cdot \epsilon_{a}\right) + S^{(1)}_{a}\left(\epsilon_{a}\cdot k_{b}\right)\right](-1)^{n}\Psi^{a}_{n+a}\right)
\end{eqnarray}
The action of $S^{(1)}_{b}$ on the momentum part and polarization part are given by

\begin{eqnarray}
 S_{b}^{(1)}(p)k_{b}^{\beta} & = & \frac{\epsilon_{n+1,\; \alpha\nu}k_{b}^{\alpha}p_{\mu}}{p\cdot k_{b}}k_{b}^{[\mu}\frac{\partial k_{b}^{\beta}}{\partial k_{b,\nu]}}\nonumber\\
  S_{b}^{(1)}(p)\epsilon_{b}^{\beta} & = & \frac{\epsilon_{n+1,\; \alpha\nu}k_{b}^{\alpha}p_{\mu}}{p\cdot k_{b}}\left(\eta^{\nu\beta}\delta^{\mu}_{\sigma}-\eta^{\mu\beta}\delta^{\nu}_{\sigma}\right)\epsilon^{\sigma}_{b}
\end{eqnarray}
The gauge fixing conditions reduce Eq.(\ref{S1-In}) to the first term of Eq.(\ref{S1-integrand}). The other term can be calculated in analogous way.

\bibliography{ref}

\providecommand{\href}[2]{#2}\begingroup\raggedright\begin{thebibliography}{10}

\bibitem{Klose}
T.~Klose, T.~McLoughlin, D.~Nandan, J.~Plefka, and G.~Travaglini,
  ``{Double-Soft Limits of Gluons and Gravitons},''
  \href{http://dx.doi.org/10.1007/JHEP07(2015)135}{{\em JHEP} {\bfseries 07}
  (2015) 135}, \href{http://arxiv.org/abs/1504.05558}{{\ttfamily
  arXiv:1504.05558 [hep-th]}}.

\bibitem{PRL}
F.~Cachazo, S.~He, and E.~Y. Yuan, ``{Scattering of Massless Particles in
  Arbitrary Dimensions},''
  \href{http://dx.doi.org/10.1103/PhysRevLett.113.171601}{{\em Phys. Rev.
  Lett.} {\bfseries 113} no.~17, (2014) 171601},
\href{http://arxiv.org/abs/1307.2199}{{\ttfamily arXiv:1307.2199 [hep-th]}}.

\bibitem{Cachazo:KLT}
F.~Cachazo, S.~He, and E.~Y. Yuan, ``{Scattering equations and
  Kawai-Lewellen-Tye orthogonality},''
  \href{http://dx.doi.org/10.1103/PhysRevD.90.065001}{{\em Phys. Rev.}
  {\bfseries D90} no.~6, (2014) 065001},
\href{http://arxiv.org/abs/1306.6575}{{\ttfamily arXiv:1306.6575 [hep-th]}}.

\bibitem{Cachazo:graviton}
F.~Cachazo, S.~He, and E.~Y. Yuan, ``{Scattering of Massless Particles:
  Scalars, Gluons and Gravitons},''
  \href{http://dx.doi.org/10.1007/JHEP07(2014)033}{{\em JHEP} {\bfseries 07}
  (2014) 033},
\href{http://arxiv.org/abs/1309.0885}{{\ttfamily arXiv:1309.0885 [hep-th]}}.

\bibitem{Cachazo:YM}
F.~Cachazo, S.~He, and E.~Y. Yuan, ``{Scattering Equations and Matrices: From
  Einstein To Yang-Mills, DBI and NLSM},''
  \href{http://dx.doi.org/10.1007/JHEP07(2015)149}{{\em JHEP} {\bfseries 07}
  (2015) 149},
\href{http://arxiv.org/abs/1412.3479}{{\ttfamily arXiv:1412.3479 [hep-th]}}.

\bibitem{Yuan}
Y.~Yuan, {\em {Scattering Equations \& S-Matrices}}.
\newblock PhD thesis, U. Waterloo (main).
\newblock
\url{https://uwspace.uwaterloo.ca/bitstream/handle/10012/9451/Ye_Yuan.pdf?sequence=3}.
\newblock

\bibitem{GeyerPRL}
Y.~Geyer, L.~Mason, R.~Monteiro, and P.~Tourkine, ``{Loop Integrands for
  Scattering Amplitudes from the Riemann Sphere},''
  \href{http://dx.doi.org/10.1103/PhysRevLett.115.121603}{{\em Phys. Rev.
  Lett.} {\bfseries 115} no.~12, (2015) 121603},
\href{http://arxiv.org/abs/1507.00321}{{\ttfamily arXiv:1507.00321 [hep-th]}}.

\bibitem{Geyer:oneloop}
Y.~Geyer, L.~Mason, R.~Monteiro, and P.~Tourkine, ``{One-loop amplitudes on the
  Riemann sphere},'' \href{http://dx.doi.org/10.1007/JHEP03(2016)114}{{\em
  JHEP} {\bfseries 03} (2016) 114},
\href{http://arxiv.org/abs/1511.06315}{{\ttfamily arXiv:1511.06315 [hep-th]}}.

\bibitem{He:loop}
S.~He and E.~Y. Yuan, ``{One-loop Scattering Equations and Amplitudes from
  Forward Limit},'' \href{http://dx.doi.org/10.1103/PhysRevD.92.105004}{{\em
  Phys. Rev.} {\bfseries D92} no.~10, (2015) 105004},
\href{http://arxiv.org/abs/1508.06027}{{\ttfamily arXiv:1508.06027 [hep-th]}}.

\bibitem{CHY:loop}
F.~Cachazo, S.~He, and E.~Y. Yuan, ``{One-Loop Corrections from Higher
  Dimensional Tree Amplitudes},''
  \href{http://dx.doi.org/10.1007/JHEP08(2016)008}{{\em JHEP} {\bfseries 08}
  (2016) 008},
\href{http://arxiv.org/abs/1512.05001}{{\ttfamily arXiv:1512.05001 [hep-th]}}.

\bibitem{Geyer:twoloop}
Y.~Geyer, L.~Mason, R.~Monteiro, and P.~Tourkine, ``{Two-Loop Scattering
  Amplitudes from the Riemann Sphere},''
  \href{http://dx.doi.org/10.1103/PhysRevD.94.125029}{{\em Phys. Rev.}
  {\bfseries D94} no.~12, (2016) 125029},
\href{http://arxiv.org/abs/1607.08887}{{\ttfamily arXiv:1607.08887 [hep-th]}}.

\bibitem{Cachazo:softtheoryext}
F.~Cachazo, P.~Cha, and S.~Mizera, ``{Extensions of Theories from Soft
  Limits},'' \href{http://dx.doi.org/10.1007/JHEP06(2016)170}{{\em JHEP}
  {\bfseries 06} (2016) 170},
\href{http://arxiv.org/abs/1604.03893}{{\ttfamily arXiv:1604.03893 [hep-th]}}.

\bibitem{DoubleSoftPRD}
F.~Cachazo, S.~He, and E.~Y. Yuan, ``{New Double Soft Emission Theorems},''
  \href{http://dx.doi.org/10.1103/PhysRevD.92.065030}{{\em Phys. Rev.}
  {\bfseries D92} no.~6, (2015) 065030},
\href{http://arxiv.org/abs/1503.04816}{{\ttfamily arXiv:1503.04816 [hep-th]}}.

\bibitem{Schwab}
B.~U.~W. Schwab and A.~Volovich, ``{Subleading Soft Theorem in Arbitrary
  Dimensions from Scattering Equations},''
  \href{http://dx.doi.org/10.1103/PhysRevLett.113.101601}{{\em Phys. Rev.
  Lett.} {\bfseries 113} no.~10, (2014) 101601},
\href{http://arxiv.org/abs/1404.7749}{{\ttfamily arXiv:1404.7749 [hep-th]}}.

\bibitem{Afkhami}
N.~Afkhami-Jeddi, ``{Soft Graviton Theorem in Arbitrary Dimensions},''
\href{http://arxiv.org/abs/1405.3533}{{\ttfamily arXiv:1405.3533 [hep-th]}}.

\bibitem{Zlotnikov}
M.~Zlotnikov, ``{Sub-sub-leading soft-graviton theorem in arbitrary
  dimension},'' \href{http://dx.doi.org/10.1007/JHEP10(2014)148}{{\em JHEP}
  {\bfseries 10} (2014) 148},
\href{http://arxiv.org/abs/1407.5936}{{\ttfamily arXiv:1407.5936 [hep-th]}}.

\bibitem{Kalousios}
C.~Kalousios and F.~Rojas, ``{Next to subleading soft-graviton theorem in
  arbitrary dimensions},''
  \href{http://dx.doi.org/10.1007/JHEP01(2015)107}{{\em JHEP} {\bfseries 01}
  (2015) 107},
\href{http://arxiv.org/abs/1407.5982}{{\ttfamily arXiv:1407.5982 [hep-th]}}.

\bibitem{CachazoStrominger}
F.~Cachazo and A.~Strominger, ``{Evidence for a New Soft Graviton Theorem},''
\href{http://arxiv.org/abs/1404.4091}{{\ttfamily arXiv:1404.4091 [hep-th]}}.

\bibitem{White}
C.~D. White, ``{Factorization Properties of Soft Graviton Amplitudes},''
  \href{http://dx.doi.org/10.1007/JHEP05(2011)060}{{\em JHEP} {\bfseries 05}
  (2011) 060},
\href{http://arxiv.org/abs/1103.2981}{{\ttfamily arXiv:1103.2981 [hep-th]}}.

\bibitem{Broedel}
J.~Broedel, M.~de~Leeuw, J.~Plefka, and M.~Rosso, ``{Constraining subleading
  soft gluon and graviton theorems},''
  \href{http://dx.doi.org/10.1103/PhysRevD.90.065024}{{\em Phys. Rev.}
  {\bfseries D90} no.~6, (2014) 065024},
\href{http://arxiv.org/abs/1406.6574}{{\ttfamily arXiv:1406.6574 [hep-th]}}.

\bibitem{Saha}
A.~P. Saha, ``{Double Soft Theorem for Perturbative Gravity},''
  \href{http://dx.doi.org/10.1007/JHEP09(2016)165}{{\em JHEP} {\bfseries 09}
  (2016) 165},
\href{http://arxiv.org/abs/1607.02700}{{\ttfamily arXiv:1607.02700 [hep-th]}}.

\bibitem{VolovichZlotnikov}
A.~Volovich, C.~Wen, and M.~Zlotnikov, ``{Double Soft Theorems in Gauge and
  String Theories},'' \href{http://dx.doi.org/10.1007/JHEP07(2015)095}{{\em
  JHEP} {\bfseries 07} (2015) 095},
\href{http://arxiv.org/abs/1504.05559}{{\ttfamily arXiv:1504.05559 [hep-th]}}.

\bibitem{Cachazo:supp}
F.~Cachazo, S.~He, and E.~Y. Yuan, ``Supplementary note for new double soft
  emission theorems,''.
  \url{https://ellisyeyuan.files.wordpress.com/2015/03/double-soft-theorems-supplementary-v12.pdf}.

\bibitem{Weinberg}
S.~Weinberg, ``{Infrared photons and gravitons},''
\href{http://dx.doi.org/10.1103/PhysRev.140.B516}{{\em Phys. Rev.} {\bfseries
  140} (1965) B516--B524}.

\bibitem{Bern:loopcor}
Z.~Bern, S.~Davies, and J.~Nohle, ``{On Loop Corrections to Subleading Soft
  Behavior of Gluons and Gravitons},''
  \href{http://dx.doi.org/10.1103/PhysRevD.90.085015}{{\em Phys. Rev.}
  {\bfseries D90} no.~8, (2014) 085015},
\href{http://arxiv.org/abs/1405.1015}{{\ttfamily arXiv:1405.1015 [hep-th]}}.

\bibitem{He:loopcor}
S.~He, Y.-t. Huang, and C.~Wen, ``{Loop Corrections to Soft Theorems in Gauge
  Theories and Gravity},''
  \href{http://dx.doi.org/10.1007/JHEP12(2014)115}{{\em JHEP} {\bfseries 12}
  (2014) 115},
\href{http://arxiv.org/abs/1405.1410}{{\ttfamily arXiv:1405.1410 [hep-th]}}.

\bibitem{Bianchi:loops}
M.~Bianchi, S.~He, Y.-t. Huang, and C.~Wen, ``{More on Soft Theorems: Trees,
  Loops and Strings},''
  \href{http://dx.doi.org/10.1103/PhysRevD.92.065022}{{\em Phys. Rev.}
  {\bfseries D92} no.~6, (2015) 065022},
\href{http://arxiv.org/abs/1406.5155}{{\ttfamily arXiv:1406.5155 [hep-th]}}.

\bibitem{Strominger:ST}
T.~He, V.~Lysov, P.~Mitra, and A.~Strominger, ``{BMS supertranslations and
  Weinberg’s soft graviton theorem},''
  \href{http://dx.doi.org/10.1007/JHEP05(2015)151}{{\em JHEP} {\bfseries 05}
  (2015) 151},
\href{http://arxiv.org/abs/1401.7026}{{\ttfamily arXiv:1401.7026 [hep-th]}}.

\bibitem{Strominger:Scat}
A.~Strominger, ``{On BMS Invariance of Gravitational Scattering},''
  \href{http://dx.doi.org/10.1007/JHEP07(2014)152}{{\em JHEP} {\bfseries 07}
  (2014) 152},
\href{http://arxiv.org/abs/1312.2229}{{\ttfamily arXiv:1312.2229 [hep-th]}}.

\bibitem{Strominger:2014pwa}
A.~Strominger and A.~Zhiboedov, ``{Gravitational Memory, BMS Supertranslations
  and Soft Theorems},'' \href{http://dx.doi.org/10.1007/JHEP01(2016)086}{{\em
  JHEP} {\bfseries 01} (2016) 086},
\href{http://arxiv.org/abs/1411.5745}{{\ttfamily arXiv:1411.5745 [hep-th]}}.

\bibitem{Campiglia:2015yka}
M.~Campiglia and A.~Laddha, ``{New symmetries for the Gravitational
  S-matrix},'' \href{http://dx.doi.org/10.1007/JHEP04(2015)076}{{\em JHEP}
  {\bfseries 04} (2015) 076},
\href{http://arxiv.org/abs/1502.02318}{{\ttfamily arXiv:1502.02318 [hep-th]}}.

\bibitem{Campiglia:2014yka}
M.~Campiglia and A.~Laddha, ``{Asymptotic symmetries and subleading soft
  graviton theorem},'' \href{http://dx.doi.org/10.1103/PhysRevD.90.124028}{{\em
  Phys. Rev.} {\bfseries D90} no.~12, (2014) 124028},
\href{http://arxiv.org/abs/1408.2228}{{\ttfamily arXiv:1408.2228 [hep-th]}}.

\bibitem{Campiglia:2016jdj}
M.~Campiglia and A.~Laddha, ``{Sub-subleading soft gravitons: New symmetries of
  quantum gravity?},''
\href{http://arxiv.org/abs/1605.09094}{{\ttfamily arXiv:1605.09094 [gr-qc]}}.

\bibitem{Campiglia:2015kxa}
M.~Campiglia and A.~Laddha, ``{Asymptotic symmetries of gravity and soft
  theorems for massive particles},''
  \href{http://dx.doi.org/10.1007/JHEP12(2015)094}{{\em JHEP} {\bfseries 12}
  (2015) 094},
\href{http://arxiv.org/abs/1509.01406}{{\ttfamily arXiv:1509.01406 [hep-th]}}.

\bibitem{He:2015zea}
T.~He, P.~Mitra, and A.~Strominger, ``{2D Kac-Moody Symmetry of 4D Yang-Mills
  Theory},'' \href{http://dx.doi.org/10.1007/JHEP10(2016)137}{{\em JHEP}
  {\bfseries 10} (2016) 137},
\href{http://arxiv.org/abs/1503.02663}{{\ttfamily arXiv:1503.02663 [hep-th]}}.

\bibitem{McLoughlin:2016uwa}
T.~McLoughlin and D.~Nandan, ``{Multi-Soft gluon limits and extended current
  algebras at null-infinity},''
\href{http://arxiv.org/abs/1610.03841}{{\ttfamily arXiv:1610.03841 [hep-th]}}.

\end{thebibliography}\endgroup
\bibliographystyle{utphys} 

\end{document}